\documentclass[amsmath,amssymb,12pt]{revtex4}
\usepackage{epsfig}
\usepackage{graphicx}
\usepackage{setspace}
\usepackage{subfigure}

\def\be{\begin{equation}}
\def\ee{\end{equation}}
\def\barr{\begin{array}}
\def\earr{\end{array}}
\def\bea{\begin{eqnarray}}
\def\eea{\end{eqnarray}}
\def\ba{\begin{eqnarray}}
\def\ea{\end{eqnarray}}

\def\sl2{$SL_2({\mathbb R})$}

\def\be{\begin{equation}}
\def\ee{\end{equation}}
\def\bea{\begin{eqnarray}}
\def\eea{\end{eqnarray}}

\def\sF{{{\rm F}\!\!\!\!\hskip.8pt\hbox{\raise1pt\hbox{/}}\,}}

\def\a{\alpha}
\def\b{\beta}

\def\o{\omega}

\def\r{\rho}

\def\t{\tau}

\def\O{\Omega}

\begin{document}

\preprint{hep-th/yymmnnn}
\title{Baryon Mass in medium  with Holographic QCD}
\author{Yunseok Seo}\thanks{{\tt yseo@hanyang.ac.kr}}\author{
Sang-Jin Sin}\thanks{{\tt sjsin@hanyang.ac.kr}}\vskip 0.5cm \affiliation{Department of physics,
BK21 Program Division, Hanyang University, Seoul 133-791, Korea }
%\date{\today}

\begin{abstract}
\vskip 2cm

\centerline{ABSTRACT}
We study the  baryon vertex (BV) in the presence of medium
using DBI action and the force balance condition  between BV  and the probe branes.
We note that a stable  BV configuration exists only  in some of the confining backgrounds.
For the system of finite density,  the issue is whether there is a canonical definition for the
baryon mass in the medium. In this work, we define it  as the energy of the deformed BV satisfying the force balance condition (FBC) with the probe brane. With FBC, lengths of the strings attached to the BV tend to be zero while the compact branes are enlongated to mimic the string. We  attribute the  deformation energy of the probe brane  to the baryon-baryon interaction. We show that for a system with heavy quarks the  baryon mass drops monotonically as a function of density
while it has minimum  in case of light quark system.
\end{abstract}

%Use showkeys class option if keyword
\maketitle
\newpage
\section{Introduction}
Recently there has been much interest in studying QCD in terms of AdS/CFT correspondence \cite{ads/cft}.
One particularly interesting question is the study of nucleon in the dense matter system.
Baryons in AdS/CFT was first introduced by Witten in \cite{wittenbaryon}, where  baryon is identified with a compact D-brane wrapping the  directions transverse  to the    $N_c$ color D-branes with $N_c$ strings attached to it.
The baryon charge is carried by the end point of the strings connecting color and flavor branes.
After the chiral model of  Sakai-Sugimoto\cite{SS1,SS2} was developed,  baryon was discussed
as the instanton on the probe branes \cite{suganuma, HRYY, SS3,HRYY2} and effective field theory of baryon was developed in \cite{HIY} in the context of phenomenological AdS/QCD model \cite{EKSS}.

The baryon density problem was first discussed in \cite{KSZ,HT} where chemical potential is identified as the asymptotic value of the electric potential on the flavor brane.
The D3/D7 system of finite temperature and finite baryon density was analyzed in \cite{NSSY} where both
Minkowski embedding as well as the black hole embedding were considered. However,
it has been argued \cite{KMMMT} that in the presence of the quark density
  the D7 brane has to touch the black hole horizon because the strings connecting D7 and the horizon can be replaced by the spiky deformation of the D7 brane.
 So it is an interesting to ask what happens if there are  baryons not just quarks.\par
 In the presence of baryon vertex (outside the black-hole horizon), the string would not touch the horizon
and there would be no reason for the spiky D-brane to touch the horizon. See figure 1-a.
The issue is then whether there is a baryon vertex \cite{wittenbaryon} for the given background and given probe branes.
The probe branes  are pulling the baryon vertex  to sustain it from the gravitational pulling of the core,  and it is not obvious  why a  compact D-brane does/doesn't exist in such background a priori.  See figure 1b.
If  a BV configuration does not exist in the black-hole background, it is not clear
whether it exist even at the confining background.
 This question is even more compelling since in D4-D8$\bar D8$ setup, a baryon vertex can exist regardless
 whether D4 background is in the confining or deconfining phase \cite{Bergman}.
\begin{figure}[!ht]
\begin{center}
{\includegraphics[angle=0, width=0.6\textwidth]{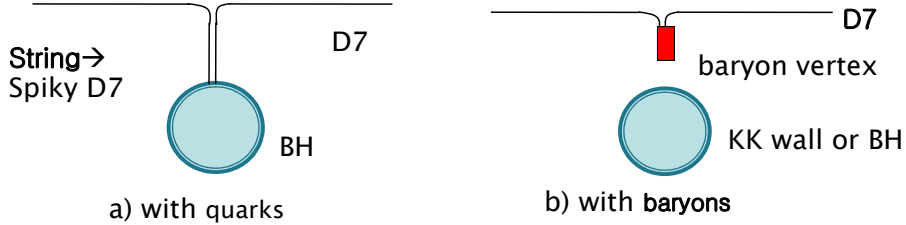}}
\caption{ Black hole embedding in quark  phase v.s Minkowski Embeddings in baryon  phase \label{fig:fig1}
}
\end{center}
\end{figure}
One of the purpose of this paper  is to examine  this issue in  various gravity backgrounds with  the probe probe branes
 using the equation of motion and  Force Balancing Condition (FBC).
 We will find that  there exist baryon vertex (BV) only in some of the confining backgrounds, while  no such object  in any of  the black hole backgrounds.\par
In case there are baryons, interesting question is the density dependence of the baryon mass when baryon vertex exists.   Here the central issue is whether there is any canonical definition of mass in the absence of Lorentz symmetry.
In  field theory, to separate the
baryon mass in the medium from the inter-baryon interaction energy has ambiguity.
However,  in the BV picture, one can define the
baryon mass in the medium simply as the  energy of deformed BV  and  we can attribute the deformation of the the probe brane  to the Baryon-Baryon interaction.
We show numerically that mass drops as density increases  for baryons with heavy quarks,
while mass v.s density profile has a minimum  for baryons with light quarks. The latter behavior is shared by the systems  D4/D6 and D4/D8$\bar D$8.

The rest of the paper goes as follows.
 In section 2, we  study  numerically to show that in the black hole background, there is no compact D-brane
 as a solution of a DBI action.
In section 3, we show that there are solutions for the baryon vertices in some confining backgrounds and study the density dependence of the baryon mass. Here we give a full details for the D4/D6 system and leave other cases
to the appendix. We conclude in section 4.

\section{Absence of baryon vertex in the  deconfining background}
 In this section, we consider D6 probe branes embedded in the non-extremal black D4 background
 and  show that the solution of a DBI action for  the baryon vertex  satisfying the force balance condition does not exist.
We discuss D3/D7 system  briefly in the Appendix (Surprisingly, the result is not completely parallel).
We start from
the gravity  background describing the deconfined phase in Euclidean signature with imaginary potential is given by
\begin{eqnarray}
ds^{2}
&=&\left(\frac{U }{R }\right)^{3/2}\left(f(U) dt^{2} +d\vec{x}^2 +dx_4^2  \right)
+\left(\frac{R}{U }\right)^{3/2}\left( \frac{dU^2}{ f(U)} +U^2 d\Omega_4^2\right) \cr
e^\phi&=&g_s\left(\frac{U }{R }\right)^{3/4},\quad F_4 =\frac{2\pi N_c}{\Omega_4}\epsilon_4, \;\; f(U)= 1-\Big(\frac{U_{0}}{U}\Big)^{3}, \;\; R^3=\pi g_s N_c l_s^3.
\label{bhbg}
\end{eqnarray}
Here  we compactify the time $t\sim t+\b_t$ to consider thermal system and
we also compactify $x_4$ for the dimensionality by  $x_4 \sim x_4 +\b_4$ and give antiperiodic boundary condition for the fermions to break the SUSY completely.
There is a horizon at $U=U_0$, and the Hawking temperature is given by
\be
T =\frac{1}{\beta_\t} =\frac{U_0}{\pi R^2},~~~~~~f(U) = 1-\frac{U_0^3}{U^3}.
\ee
Introducing a dimensionless coordinate $\xi$  by $\frac{d\xi^2}{\xi^2}=\frac{dU^2}{U^2f(U)}$,
the background geometry becomes
\be\label{d4bhmetric}
ds^2 = \left(\frac{U }{R }\right)^{3/2}\left(f(U) dt^2 +d\vec{x}^2 + dx_4^{2} \right)
+\left(\frac{R}{U }\right)^{3/2}\left(\frac{U}{\xi}\right)^2\left(d\xi^2 +\xi^2 d\Omega_4^2\right),
\ee
where $U$ and $\xi$ are related by
\be
\left(\frac{U}{U_{0}}\right)^{3/2} = \frac{1}{2}\left(\xi^{3/2}+\frac{1}{\xi^{3/2}}\right), \;\; {\rm and } \;\;
f = \left(\frac{1-\xi^{-3}}{1+\xi^{-3}}\right)^2 \equiv \frac{\omega_{-}^2}{\omega_{+}^2}.
\ee

A baryon in the three-dimensional theory corresponds to the D4 brane wrapped on an $S^4$
on which $N_c$ fundamental strings terminate. In this configuration, the background four-form field strength can
couple to the world volume gauge field $A_{(1)}$ via Chern-Simons term.\par
The  metric (\ref{d4bhmetric}) can be written as
\be
ds^2 = \left(\frac{U}{R}\right)^{3/2}\left(f dt^2 +d\vec{x}^2 +dx_4^2  \right)
+R^{3/2}\sqrt{U}\left(\frac{d\xi^2}{\xi^2} +d\theta^2 +\sin^2\theta d\Omega_{3}^{2}\right),
\ee
We take $(t,\theta_{\a})$ as  world volume coordinates of a compact D4 brane, and turn on the $U(1)$ gauge field on it to have $F_{t\theta}\ne 0$.
 As the ansatz for the embedding of compact D4, we assume the $SO(4)$ symmetry so that position of D4 brane and gauge field depend only on $\theta$ i.e. $\xi=\xi(\theta)$, $A_{t}=A_t(\theta)$, where $\theta$  measure the polar angle of $S^4$ from the north pole. The induced metric on D4 brane is
\be
ds_{D4}^2 = \left(\frac{U}{R}\right)^{3/2}f(U)dt^2
+R^{3/2}\sqrt{U}\left[\left(1+\frac{\xi'^2}{\xi^2}\right)d\theta^2 +\sin^2\theta d\Omega_{3}^{2}\right],
\ee
%The difference between confinement phase is that there is another $f$ factor in front of $dt^2$. With a same procedure
%to the previous section,
where $\xi' =\partial \xi/\partial \theta$.
The DBI action for single D4 brane with $N_c$ fundamental string can be written as
\bea\label{bary-d4}
S_{D4} &=& -\mu_4 \int e^{-\phi} \sqrt{{\rm det}(g+2\pi \alpha' F)}+\mu_4 \int  A_{(1)}\wedge G_{(4)} \cr\cr
&=& \t_4 \int dtd\theta \sin^3\theta
\left[-\sqrt{ \frac{\o_-^2}{\o_+^{2/3}} (\xi^2 +\xi'^2)-\tilde{F}^2}
+3 \tilde{A}_t \right] \cr\cr
&=& \int dt {\cal L}_{D4},
\eea
where
\bea
\t_4 &=& \mu_4 \O_3 g_s^{-1} R^{3}\frac{U_{0}}{2^{2/3}}=\frac{N_c U_{0}}{2^{8/3}(2\pi l_s^2)} \cr\cr
\tilde{F} &=& \frac{2\pi \a'2^{4/6}}{U_{0}}F_{t\theta},~~~~~~ \tilde{A}_t=\frac{2^{2/3}}{U_{0}}\cdot 2\pi\a'A_t.
\eea

\par
The dimensionless displacement can be defined as follows;
\bea
\frac{\partial {\cal L}_{D4}}{\partial \tilde{F}}
&=& \frac{\sin^3\theta \tilde{F}}{\sqrt{({\o_-^2}/{\o_+^{2/3}})(\xi^2 +\xi'^2)
-\tilde{F}^2}} \cr\cr
&\equiv& -D(\theta).
\eea
Then the equation of motion for gauge field is
\be
\partial_{\theta}D(\theta)=-3\sin^3 \theta.
\ee
Integrating, we get
\be\label{displacement}
D(\theta)=2(2\nu -1)+3(\cos\theta -\frac{1}{3}\cos^3\theta),
\ee
where the integration constant $\nu$ determines the number of fundamental sting
( $\nu N_c$ strings are attached at south pole and $(1-\nu)N_c$ strings at north pole, we set $\nu =0$)\par
\begin{figure}[!ht]
\begin{center}
\subfigure[]{\includegraphics[angle=0, width=0.3\textwidth]{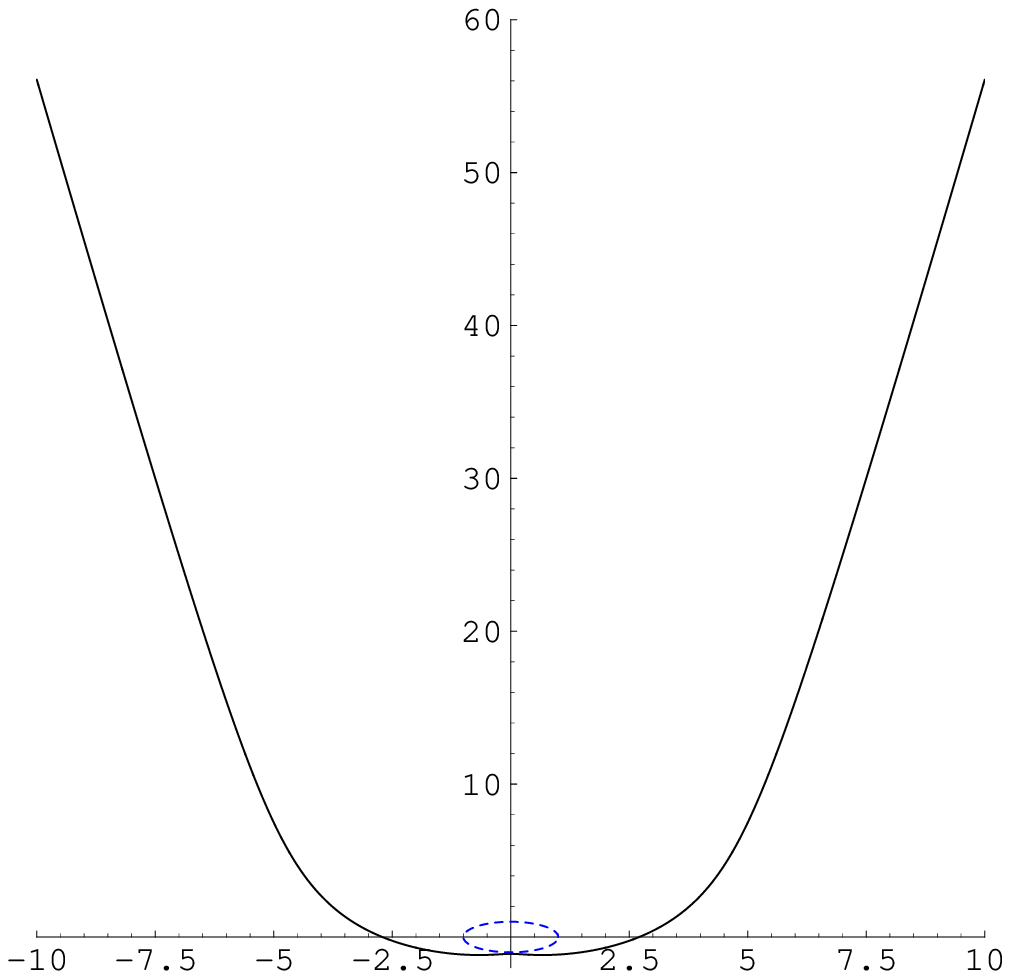}}
\subfigure[]{\includegraphics[angle=0, width=0.45\textwidth]{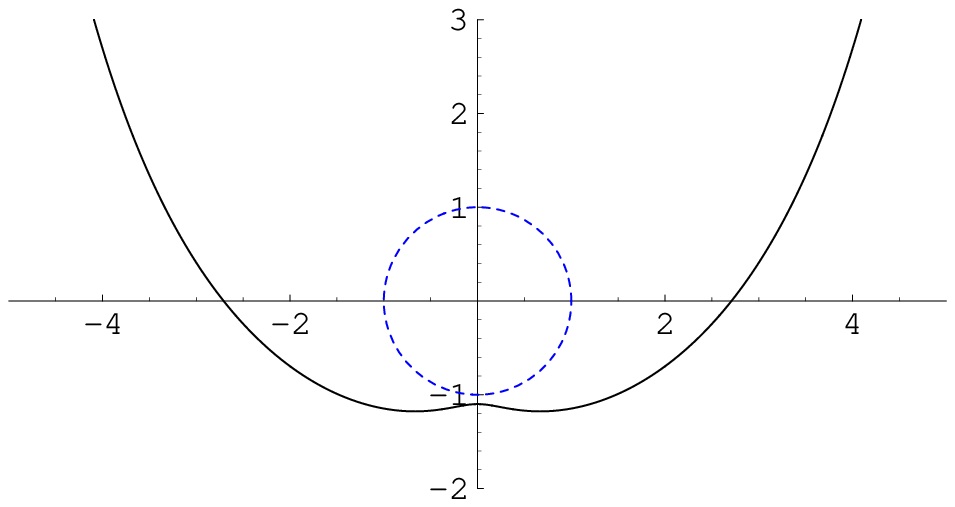}}
\caption{(a) Solution of baryon D4 brane in deconfined phase. (b) Solution near the origin\label{fig:decon-baryd4}
}
\end{center}
\end{figure}

By performing Legendre transformation, we can get `Hamiltonian'
\bea\label{d4bh}
{\cal H}_{D4} &=&\tilde{F}\frac{\partial {\cal L}_{D4}}{\partial \tilde{F}}-{\cal L}_{D4}\cr\cr
&=&\t_4 \int d\theta \sqrt{\frac{\o_-^2}{\omega_+^{2/3}} (\xi^2 +\xi'^2)}\sqrt{D(\theta)^2+ \sin^6\theta}.
\eea
The numerical solution of the equation of motion for above Hamiltonian is drawn at FIG. \ref{fig:decon-baryd4}.
There is no closed D4 brane solution which can be a baryon vertex, except a D4 brane wrapping blackhole
horizon itself. We also get same results of absence of D5 baryon vertex in D3 deconfing background.\par
Since there is no candidate for the baryon vertex, there is no point to discuss the baryon on the D6 brane theory.
All the baryon charges exists as quarks (strings attached on it) and there are no Minkowski embedding in this background.
This gravity result is consistent with that of gauge theory result in deconfined phase: there is no baryon  in deconfined phase.
The thermodynamical properties of D6 brane in the blackhole background was discussed recently in \cite{Matsuura:2007zx}, so we do not discuss them here.

\section{ The baryon mass in medium  in confining background}
%\subsection{D4/D6 system}
The non-supersymmetric geometry for confining background of D4 in Euclidean signature is given by
\begin{eqnarray}
ds^{2}
&=&\left(\frac{U }{R }\right)^{3/2}\left(\eta_{\mu\nu}dx^\mu dx^\nu + f(U) dx_4^{2} \right)
+\left(\frac{R}{U }\right)^{3/2}\left( \frac{dU^2}{ f(U)} +U^2 d\Omega_4^2\right) \cr
e^\phi&=&g_s\left(\frac{U }{R }\right)^{3/4},\quad F_4 =\frac{2\pi N_c}{\Omega_4}\epsilon_4, \;\; f(U)=
1-\Big(\frac{U_{KK}}{U}\Big)^{3}, \;\; R^3=\pi g_s N_c l_s^3.
\label{adsm}
\end{eqnarray}
This background is related to the previous one by the  double Wick rotation
\be
x_4 \longleftrightarrow t,~~~~~~~~t \longleftrightarrow x_4,~~~~~~~~~U_0 \longleftrightarrow U_{KK}
\ee
The Kaluza-Klein mass is defined as the inverse radius of the $x_4$ direction:
$M_{KK}=\frac{3}{2}\frac{U^{1/2}_{KK}}{R^{3/2}}$.
%the $g^2_{YM}$ is defined by $ {1}/{g^2_{YM}}= {\delta\tau} (2\pi \alpha')^2T_{D4}$.
While $U_{KK}, g_s, R$ are bulk parameters,
$M_{MM}$ and $g^2_{YM}$ is the parameter of the gauge theory.
They can be related by
\be
g_s=\frac{\lambda}{2\pi l_sN_c M_{KK}}, \quad U_{KK}=\frac{2}{9}\lambda M_{KK} l_s^2,
\quad R^3=\frac{\lambda l_s^2}{2M_{KK}} ,\quad \lambda=g_{YM}^{2}N_{c}.
\ee

Introducing a same dimensionless coordinate $\xi$ as previous section,
the bulk geometry becomes
\be\label{d4bgmetric}
ds^2 = \left(\frac{U }{R }\right)^{3/2}\left(dt^2 +d\vec{x}^2 + f(U) dx_4^{2} \right)
+\left(\frac{R}{U }\right)^{3/2}\left(\frac{U}{\xi}\right)^2\left(d\xi^2 +\xi^2 d\Omega_4^2\right),
\ee
and $U$, $\xi$ are related by
$\left(\frac{U}{U_{KK}}\right)^{3/2} = \frac{1}{2}\left(\xi^{3/2}+\frac{1}{\xi^{3/2}}\right).$

%\subsection{Confined Phase}
\subsection{Baryon vertex - D4 brane}
\begin{figure}[!ht]
\begin{center}
{\includegraphics[angle=0, width=0.3\textwidth]{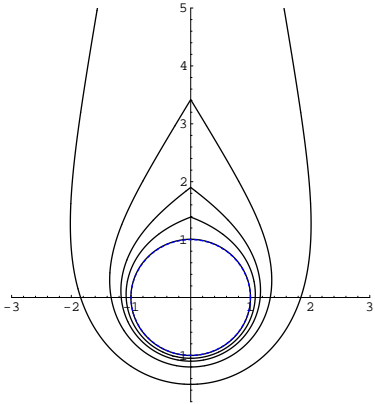}}
\caption{Shape of D4 brane for different $\xi_0$.\label{fig:baryon-d4}
}
\end{center}
\end{figure}

In confining phase, we consider same configuration of baryon D4 brane wrapping on $S^4$.
The background metric (\ref{d4bgmetric}) can be written as
\be
ds^2 = \left(\frac{U}{R}\right)^{3/2}\left(dt^2 +f dx_4^2 +d\vec{x}^2  \right)
+R^{3/2}\sqrt{U}\left(\frac{d\xi^2}{\xi^2} +d\theta^2 +\sin^2\theta d\Omega_{3}^{2}\right),
\ee
%We take $(t,\theta_{\a})$ as a world volume coordinate of D4 brane, and turn on the $U(1)$ gauge field on it,
%$F_{t\theta}\ne 0$. For simplicity, we assume that position of D4 brane and gauge field depends only on $\theta$ i.e.
%$\xi=\xi(\theta)$, $A_{t}=A_t(\theta)$, where $\theta$ is the polar angle in spherical coordinates.
We are using same world volume coordinate of D4 brane as previous section.
The induced metric on the compact D4 brane is
\be\label{d4met}
ds_{D4}^2 = \left(\frac{U}{R}\right)^{3/2}dt^2
+R^{3/2}\sqrt{U}\left[\left(1+\frac{\xi'^2}{\xi^2}\right)d\theta^2 +\sin^2\theta d\Omega_{3}^{2}\right],
\ee
and $\xi' =\partial \xi/\partial \theta$.
The DBI action for single D4 brane with $N_c$ fundamental string can be written as \cite{Callan:1999zf}
\bea\label{bary-d4}
S_{D4} &=& -\mu_4 \int e^{-\phi} \sqrt{{\rm det}(g+2\pi \alpha' F)}+\mu_4 \int  A_{(1)}\wedge G_{(4)} \cr\cr
&=& \t_4 \int dtd\theta \sin^3\theta
\left[-\sqrt{ \o_+^{4/3} (\xi^2 +\xi'^2)-\tilde{F}^2}
+3 \tilde{A}_t \right] \cr
&=& \int dt {\cal L}_{D4},
\eea
\bea
{\rm with}~~~~
t_4 &=& \mu_4 \O_3 g_s^{-1} R^{3}\frac{U_{KK}}{2^{2/3}}=\frac{N_c U_{KK}}{2^{8/3}(2\pi l_s^2)} \cr\cr
\tilde{F} &=& \frac{2\pi \a'F_{t\theta}2^{4/6}}{U_{kk}},~~~~~~ \tilde{A}_t=\frac{2^{2/3}}{U_{KK}}\cdot 2\pi\a'A_t.
\eea
The Legendre transformed `Hamiltonian' is
\bea\label{d4h}
{\cal H}_{D4} &=&\tilde{F}\frac{\partial {\cal L}_{D4}}{\partial \tilde{F}}-{\cal L}_{D4}\cr\cr
&=&\t_4 \int d\theta \sqrt{\omega_+^{4/3} (\xi^2 +\xi'^2)}\sqrt{D(\theta)^2+ \sin^6\theta},
\eea
where displacement $D(\theta)$ is same as (\ref{displacement}).
Solutions of equation of motion for above Hamiltonian can be obtained   numerically.  We set $\nu=0$
because we assume that all fundamental strings are attached at north pole($\theta=\pi$).
Then we should impose smooth boundary
condition ($\xi'(0)=0$ and $\xi(0)=\xi_0$) at $\theta=0$. Numerical solutions are
parameterized by initial value of $\xi_0$. The solutions corresponding to different $\xi_0$'s  are drawn in FIG. \ref{fig:baryon-d4}.
In this figure, $N_c$ fundamental strings are attached at the cusp. This configuration will be stable only if there
is tension balance between two object.\par
If we call the position of the cusp of D4 brane  by $U_c$, the force from the D4 brane tension
can be obtained by  varying the Hamiltonian of D4 brane with respect to $U_c$ while keeping other variables;
\bea\label{force-d4}
F_{D4}&=& \frac{\partial{{\cal H}}}{\partial U_c} \Bigg|_{fix~other~values} \cr\cr
&=& N_c T_F \left(\frac{1+\xi_c^{-3}}{1-\xi_c^{-3}}\right)
\frac{\xi_c'}{\sqrt{\xi_c^2 +\xi_c'^2}},
\eea
where $ T_F =\frac{2^{2/3}\t_4 }{N_c U_{KK}}$ is tension of fundamental string.
The tension of D4 brane is always smaller or equal to the fundamental string. Therefore, if there were no probe branes, the cusp should be pulled up to infinity  and the final configuration of D4
would be `tube-like' shape as in  \cite{callan}.
So far  we considered single  compact  D4 brane with $N_c$ fundamental strings.
Now we need to study the deformation of the D6 which is pulling the compact D4 brane by fundamental strings.

\subsection{Fundamental strings connecting baryon vertex  and the probe brane}
If we take D4-D6 system  with confining background  and consider the baryon vertex as compact D4, both D4 and D6 deform due to the
interaction between them through fundamental string (F1) and the D4 background.
When strings and D-branes are in contact, the length of the F1 tends to be  zero  while the D-branes are deformed
to replace them for minimum energy configuration. One may say that  `` F1 is the most expensive  object of all branes".
Therefore the strings connecting the D4  and D6 have zero length and the D4 and D6 are contacting at a point.
In fact there is no way to connect  D-branes of different dimensionality by a tube-like deformation  even before we consider a solution of a DBI system.  The point contact is the only possible configuration.

\subsection{Probe D6 brane}
Now, we put probe D6 brane
where the other end points of fundamental strings are attached. The string endpoints can be understood as point charges on D6 brane.  The gauge potential which couples to this point source is $A_{t}$.
The the bulk metric (\ref{d4bgmetric}) can be written as
\be
ds^2 = \left(\frac{U }{R }\right)^{3/2}\left(dt^2 +d\vec{x}^2 + f(U) dx_4^{2} \right)
+\left(\frac{R}{U }\right)^{3/2}\left(\frac{U}{\xi}\right)^2\left(d\r^2 +\r^2 d\Omega_2^2+dy^2 +y^2 d\phi^2\right),
\ee
where D6 brane world volume coordinates are $(t,\vec{x},\r,\theta_{\a})$.
The embedding ansatz is that only $y$ depends on $\r$ and  $\phi=0$. The induced metric on D6 brane is
\be
ds^2_{D6} =\left(\frac{U}{R}\right)^{3/2}(dt^2 + d\vec{x}^2) +\left(\frac{R}{U}\right)^{3/2}\left(\frac{U}{\xi}\right)^2
\left[(1+\dot{y}^2)d\r^2 +\r^2 d\O_2^2\right],
\ee
where $\dot{y}=\partial y/\partial\r$.\par
The DBI action for $N_f$ D6 brane is
\bea
S_{D6}=  \int dt {\cal L}_{D6} &=& -N_f \mu_6 \int e^{-\phi}\sqrt{{\rm det}(g+2\pi \a' F)} \cr\cr
&=& -\t_6 \int dtd\r \r^2  \o_+^{4/3} \sqrt{\o_+^{4/3} (1+\dot{y}^2)-\tilde{F}^2},
\eea
where
\bea
\t_6 =\frac{1}{4}N_f \mu_6 V_3  \O_2 g_s^{-1} U_{KK}^3 ,\quad
\tilde{F}=\frac{2\cdot2^{2/3}\pi\a'F_{t\r}}{U_{KK}}.
\eea
We define dimensionless quantity $\tilde{Q}$ from the equation of motion for $\tilde{F}$;
\be
\frac{\partial S_{D6}}{\partial \tilde{F}}
=\frac{ \r^2 \o_+^{4/3}\tilde{F}}{\sqrt{\o_+^{4/3} (1+\dot{y}^2)-\tilde{F}^2}}
\equiv \tilde{Q}.
\ee
The hamiltonian  is connected to the number of point sources (number of fundamental strings) $Q$ by
\be
\tilde{Q}=\frac{U_{KK}Q}{2\cdot2^{2/3}\pi\a'\t_6}.
\ee

\begin{figure}[!ht]
\begin{center}
\subfigure[]{\includegraphics[angle=0, width=0.45\textwidth]{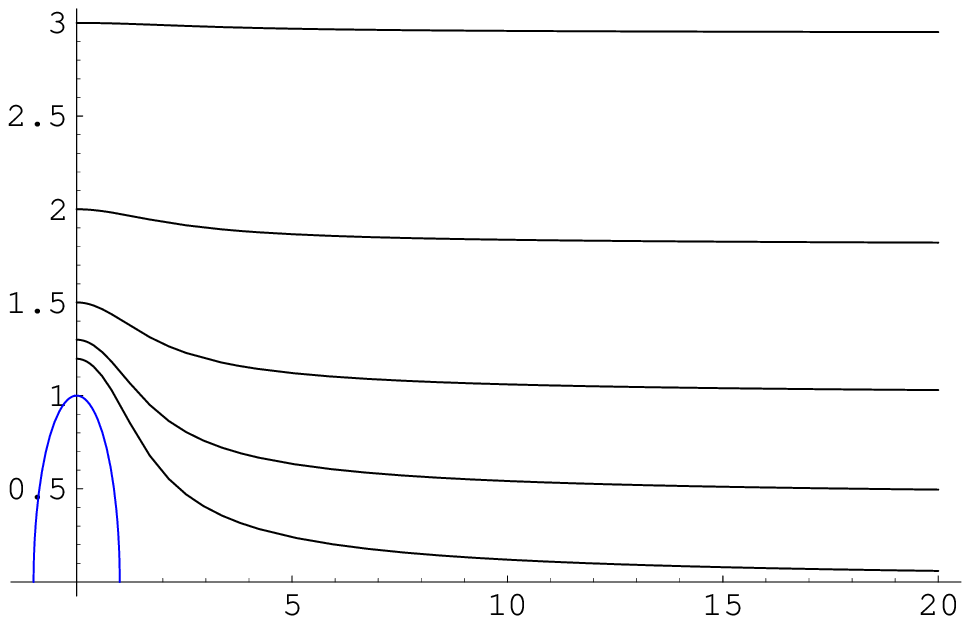}}
\subfigure[]{\includegraphics[angle=0, width=0.45\textwidth]{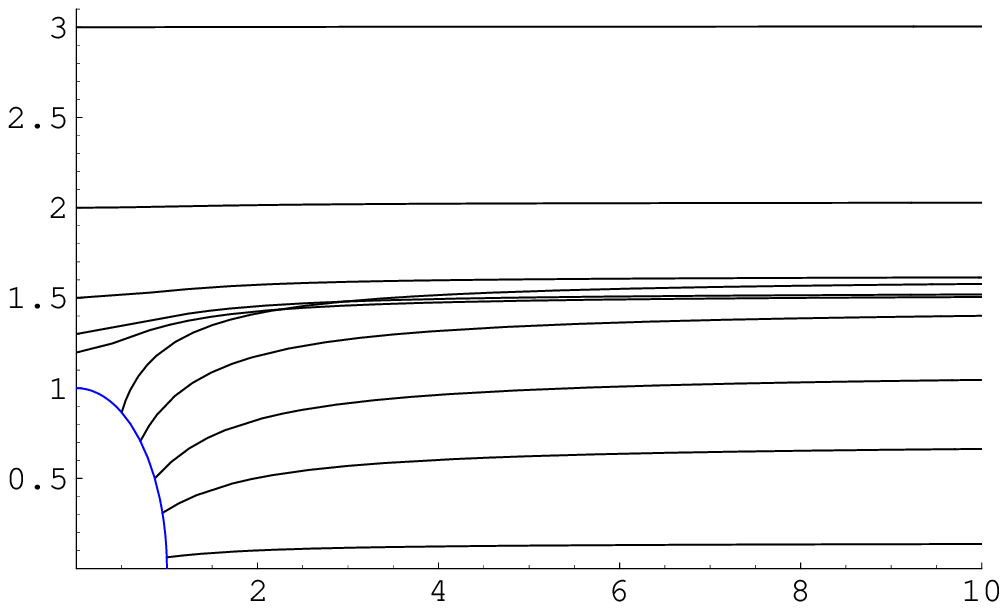}}
\caption{D6 brane embedding with different $y_0$ in the case of $Q=0$.  Blue circle denote singularity or horizon.
(a) In confinement phase. (b) In deconfined phase\label{fig:d6-01}
}
\end{center}
\end{figure}

 The `Hamiltonian' can be obtained by Legendre transformation;
\bea
{\cal H}_{D6} &=&\tilde{F}\frac{\partial S_{D6}}{\partial \tilde{F}}-S_{D6} \cr\cr
&=& \t_6 \int d\r \sqrt{\o_+^{4/3}\left(\tilde{Q}^2+\r^4 \o_+^{8/3}\right)}\sqrt{1+\dot{y}^2} \cr\cr
&=& \t_6 \int d\r V(\r)\sqrt{1+\dot{y}^2}
\eea
The equation of motion after eliminating the gauge field is written explicitly by
\be\label{emod6}
\frac{\ddot{y}}{1+\dot{y}^2}+\frac{\partial \log V}{\partial \r}\dot{y}-\frac{\partial\log V}{\partial y}=0,
\ee
and the Gauss-law constraint gives
\be
\tilde{F}=\frac{\tilde{Q}\o_+^{4/3}\sqrt{1+\dot{y}^2}}{\sqrt{\tilde{Q}^2 +\r^4 \o_+^{8/3}}}.
\ee
We can solve the eqaution (\ref{emod6}) if    boundary condition is given:
In $\tilde{Q}=0$ case, the solutions of probe D6 brane embedding are drawn in FIG. \ref{fig:d6-01}(a).
Here we impose $\dot{y}(0)=0$   for the smoothness.
The value of $y$ at the infinity corresponds to $m_q$.
In $\tilde{Q}\ne 0$ case,  there are fundamental strings which connect baryon D4 brane
with probe D6 brane. In this case,  force balance condition request that $\dot{y}(0)$ be non-zero.
The force at the cusp of probe D6 brane can be obtained as;
\bea\label{force-d6}
\hat{F}_{D6}&=&\frac{\partial {\cal H}_{D6}}{\partial U_c} \Bigg|_{\partial} \cr\cr
&=&\frac{Q}{2\pi\a'}\left(\frac{1+\xi_c^{-3}}{1-\xi_c^{-3}}\right)
\frac{\dot{y}_c}{\sqrt{1+\dot{y}_c^2}}.
\eea
The whole configuration (compact D4+F1+D6) can be stationary  if there is a force balance condition.
For $Q$ fundamental strings, the  number of baryon D4 is $Q/N_c$.
The condition  which
makes the system to be stationary ($\r=0$, $y_c =\xi_c$) is:
\be\label{force-balance2}
F_{D6}=\frac{Q}{N_c} F_{D4},
\ee
which is simplified  to be
\be
\dot{y}_c = \frac{\xi_c'}{y_c}.
\ee
\begin{figure}[!ht]
\begin{center}
\subfigure[] {\includegraphics[angle=0, width=0.4\textwidth]{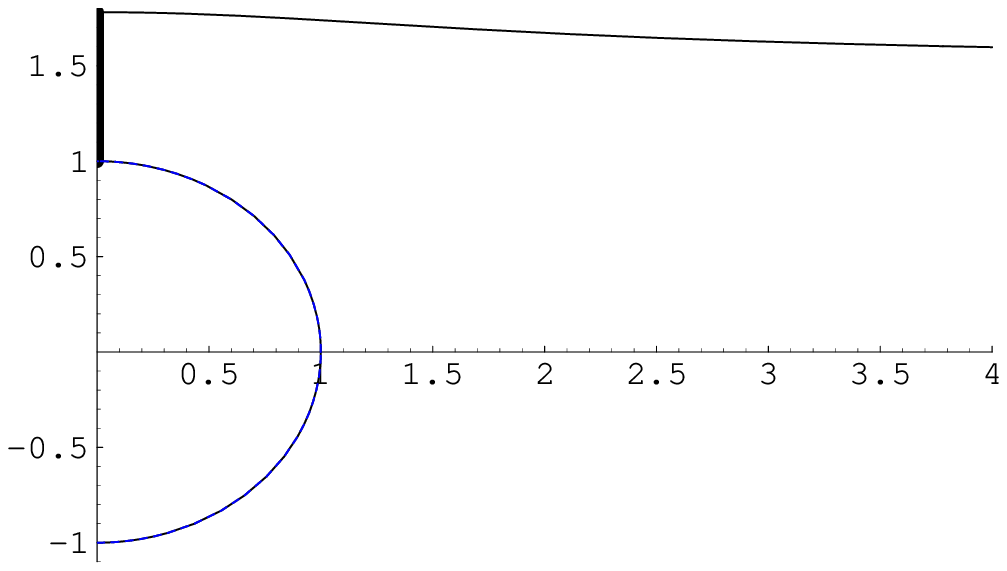}}
\subfigure[] {\includegraphics[angle=0, width=0.38\textwidth]{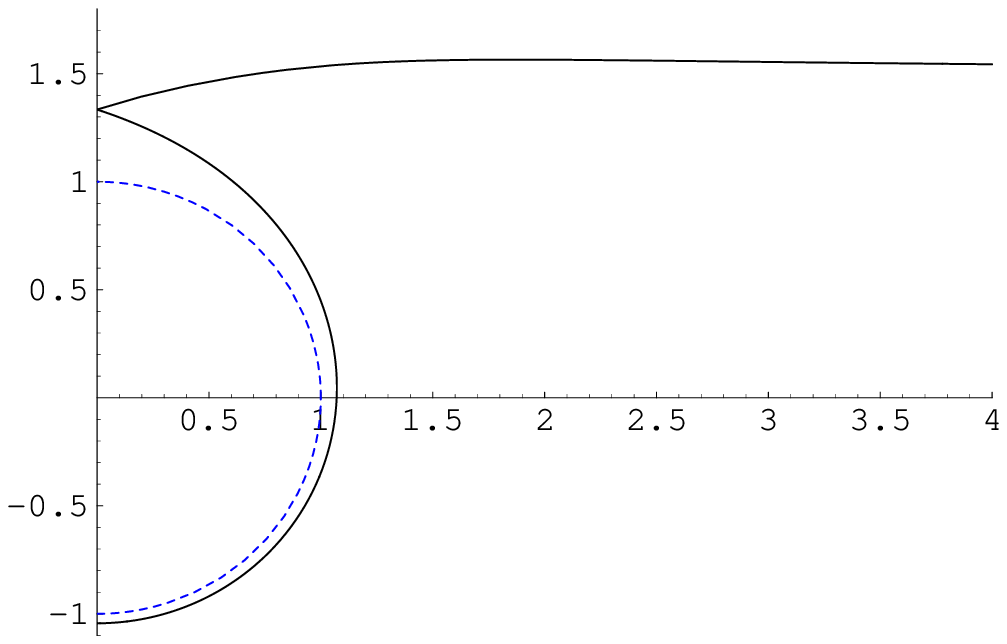}}
\caption{\label{d6forcebal}Probe D6 brane embedding for $m_q=1.5$ (a)without force balance condition.
Blue circle denote singularity where baryon D6 brane  wrapped around. Thick vertical line denotes fundamental strings
which are understood as a source. (b)with force balance condition.  }
\end{center}
\end{figure}
With this, the value of $\dot{y}_c$ which satisfy force
balance condition is uniquely determined for given value of $\xi_o$ and  $m_q$.
The difference between the behaviors of probe D6 branes with and without the force balance
condition is drawn in FIG. \ref{d6forcebal};
  In (a) we draw for the case Without FBC: $Q$ fundamental strings stretched between  baryon D4 and probe D6 brane.
  (b) is for the case with FBC: baryon D4 brane is pulled up while probe D6 brane is pulled down such that the total system is stationary.

Since  we   established the existence of the baryon vertex, we now want to study how the the mass of baryon depends on the medium density.
The difficulty comes from the broken symmetry: The very definition of the
mass is the casimir Poincare invariance which is broken in the presence of the medium
and it is not obvious what is the most natural definition of  the baryon mass inside a medium.

\begin{figure}[!ht]
\begin{center}
\subfigure[] {\includegraphics[angle=0, width=0.45\textwidth]{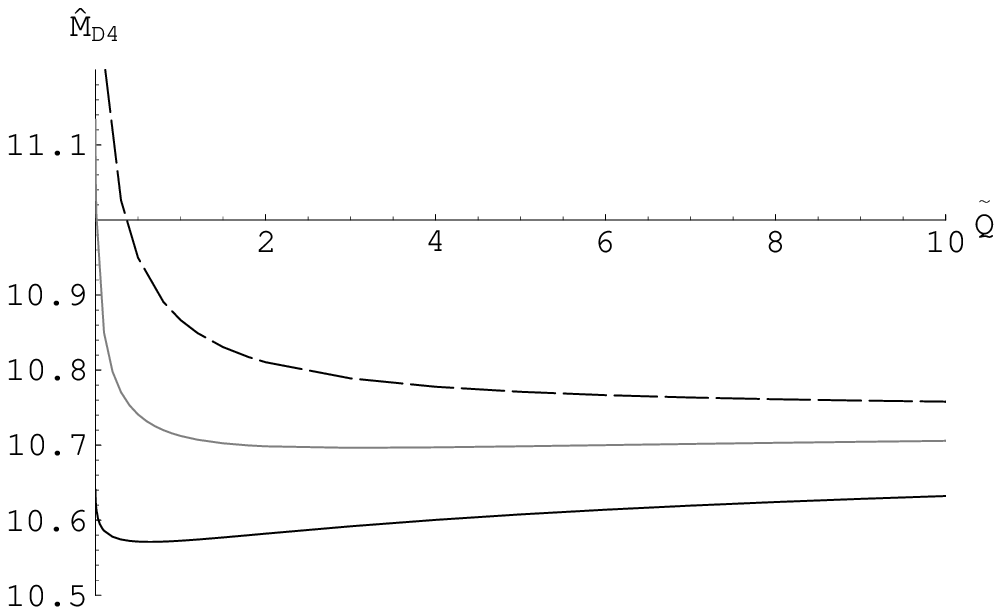}\label{fig:md4s-01}} %%&
\subfigure[] {\includegraphics[angle=0, width=0.45\textwidth]{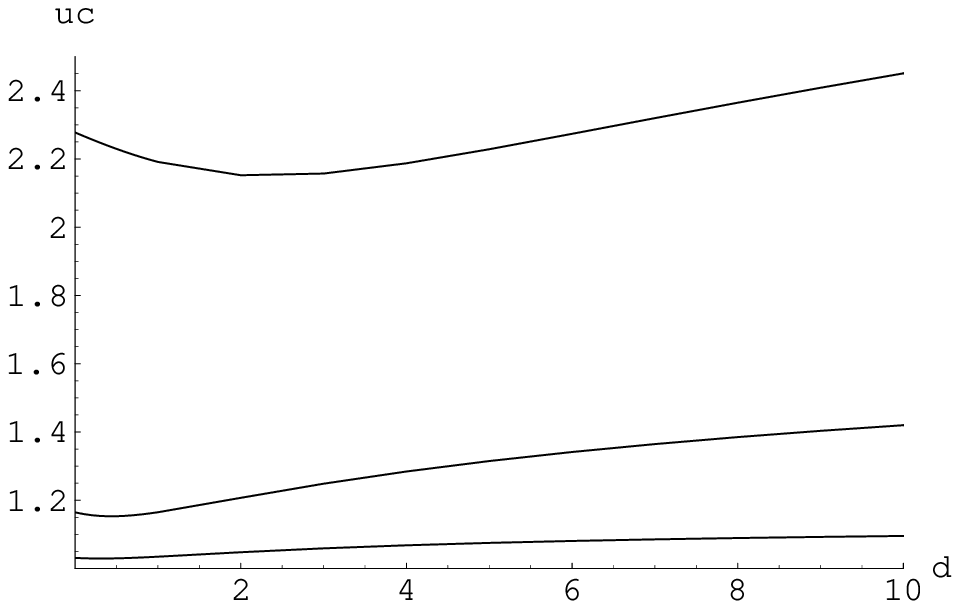}\label{fig:ucd}}
\caption{(a) Baryon Mass in medium for  D4/D6 system with current quark masses
 $m_q =1.5$(dashed line), $m_q=1$(gray line), $m_q=0$(real line).
  (b) The position (mass) of D4   brane as a function of the electric displacement d for $l=0.5$, $l=1$ and
$l=1.5$ from top to below($l=1$ result is obtained in \cite{Bergman}).}
\end{center}
\end{figure}

As we discussed before, the length of the string of the baryon vertex is zero and therefore D4 and D6 are contacting each other at a point. There are two sources of contribution of the baryon mass change: One is the deformation of the
compact D4 brane from the spherical shape and the other is the deformation of the probe D6 brane
from the zero charge configuration. We believe that the latter is responsible for the baryon-baryon interaction while
the former is related to the quark-quark interaction to form a baryon in the medium.
Therefore we define the mass in the medium as the energy of the deformed compact D-brane.
Then, the mass of a baryon is proportional to the value of $\xi_0$.
The density dependence of a mass of baryon D4 brane for several value of $m_q$ is drawn in FIG. \ref{fig:md4s-01}.
For large current quark mass, the baryon mass decreases as a function of the density. On the other hand, for the small quark mass it has a minimum. Similar behavior was observed in D4/D8/$\bar D8$ system\cite{Bergman}, where we have a zero current quark mass. See FIG. \ref{fig:ucd}.

\subsection{Binding energy of  baryons and their interaction energy}
Consider the configuration where baryon vertex  and probe brane is not connected. The position of baryon vertex is lowest possible value $U_{KK}$  and the shape is spherical.  See   Fig. \ref{d6forcebal} (a).
The energy of this configuration ${\cal F}^{(0)}$  is given by
\bea\label{F0}
{\cal F}^{(0)} &=& {\cal H}_{D6}^{(0)} + N_B {\cal H}_{D4}^{(0)} + Q {\cal H}_{F1} \cr\cr
&=& \t_6 \left[\hat{\cal H}_{D6}^{(0)}(\tilde{Q}=0) +\frac{\tilde{Q}}{4}\hat{\cal H}^{(0)}_{D4}(\tilde{Q})
 +\tilde{Q}\int_1^{\xi_c^0}(1+\xi^{-3})^{2/3}d\xi \right],
\eea
where ${\cal H}_{D6}^{(0)}$ is the probe D6 brane energy without charge,
${\cal H}_{D4}^{(0)}$ is the compact and spherical D4 brane energy without attached string and
 $\xi_c^{0}$ is a position of D6 brane at $\r=0$ and $N_B$ is a baryon number $Q/N_c$.
%The   Hamiltonian with superscript 0 is energy of the configuration without interaction.
The energy of configuration (b) in Fig. \ref{d6forcebal} with FBC imposed is given by
\be\label{FF}
{\cal F}=\t_6 \left[\hat{\cal H}_{D6}(\tilde{Q}) +\frac{\tilde{Q}}{4}\hat{\cal H}_{D4}\right].
\ee
We define the baryon mass in a medium as the energy of the deformed D4 with FBC imposed:
\be
M_B:= {\cal H}_{D4}.
\ee
The binding energy of a baryon can be defined as the energy difference between the baryon vertices before and after the deformation:
\be
E_B=    {\cal H}^{(0)}_{D4}( { Q} )
 + N_c {\cal H}_{F1} -  M_B.
\ee
 The baryon-baryon interaction energy can be defined as the deformation energy of the probe brane, that is,
 the difference between the energy of  `D6 without charge' and that of `D6 with charge and  FBC'.
 Since this  energy present for any pair of the charges we expect that it is proportional to the $Q^2 $ at least for small $\tilde Q$. Indeed the numerical result shows the quadratic dependence on the $Q$.
When this quantity is positive (negative), baryon-baryon interaction is dominated by the repulsion(attraction).

\begin{figure}[!ht]
\begin{center}
\subfigure[] {\includegraphics[angle=0, width=0.4\textwidth]{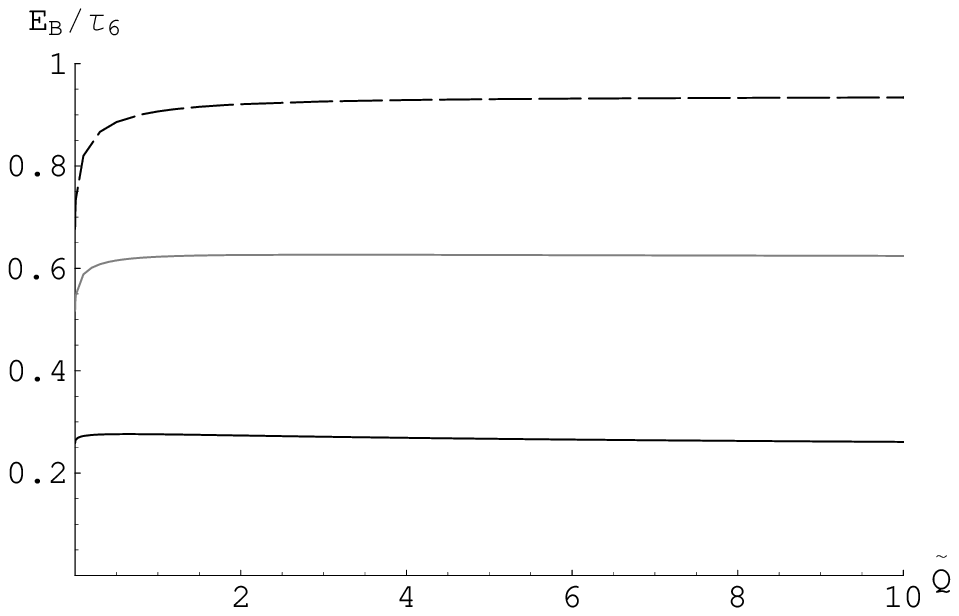}}
\subfigure[]{\includegraphics[angle=0, width=0.38\textwidth]{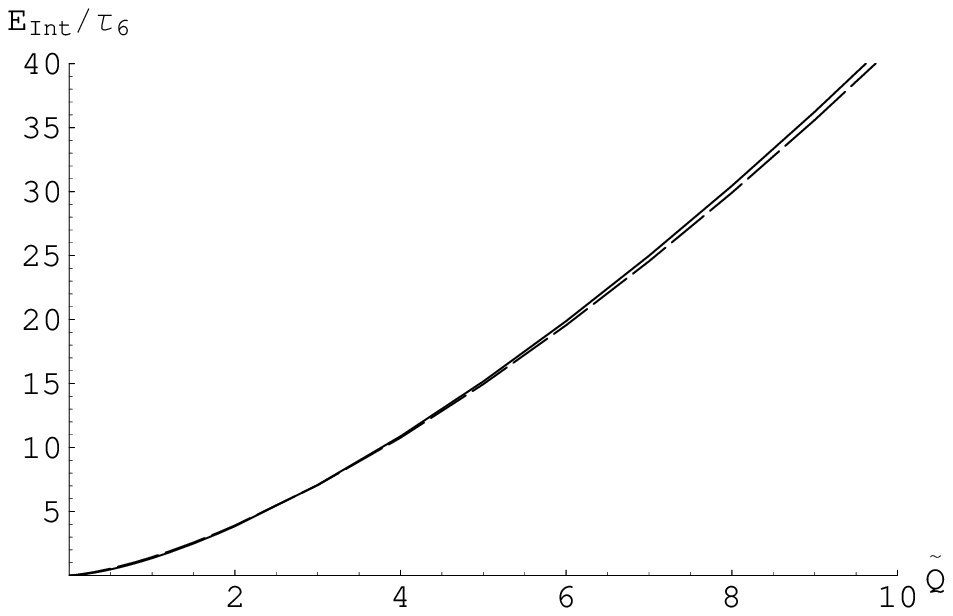}}
\caption{(a) Binding energy for
$m_q =0$(real line),$m_q=1$(gray line), $m_q=1.5$(dashed line). (b) Interaction energy
for
$m_q =0$(real line),$m_q=1.5$(dashed line)($m_q=1$ line is almost same as dashed line)\label{fig:binding} }
\end{center}
\end{figure}

The difference of  ${\cal F}$ and  ${\cal F}^{(0)}$  gives  total binding energy  plus
interaction energy of the multi-baryon system:
\be
{\cal F}-{\cal F}^{(0)} =-N_B E_B +E_{Interaction}.
\ee
Numerical result of binding energy as well as the
 interaction  energy  is drawn in FIG. \ref{fig:binding} \par

\begin{figure}[!ht]
\begin{center}
\subfigure[]{\includegraphics[angle=0, width=0.45\textwidth]{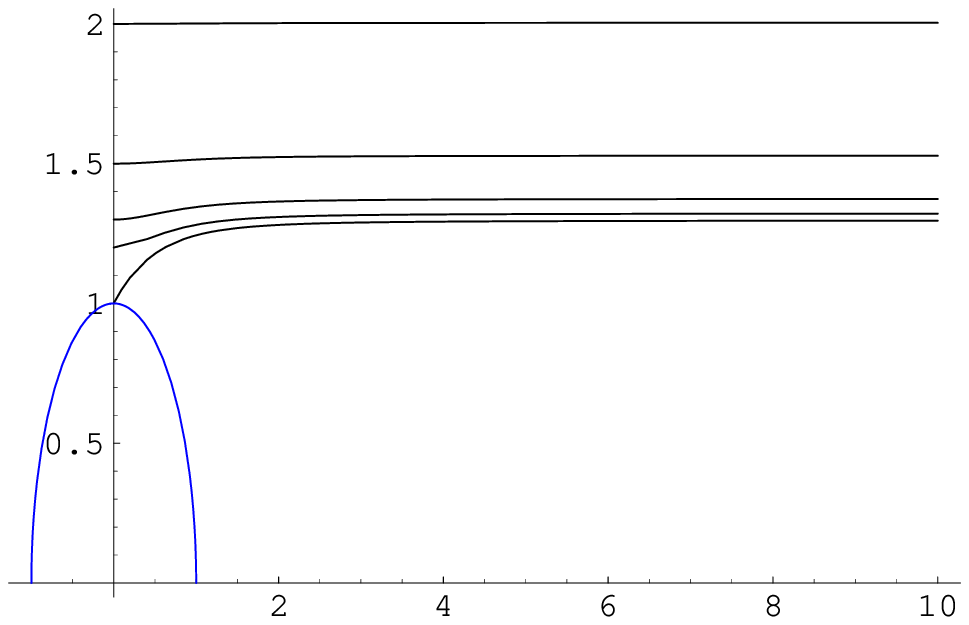}\label{fig:d7I-01}}
\subfigure[]{\includegraphics[angle=0, width=0.45\textwidth]{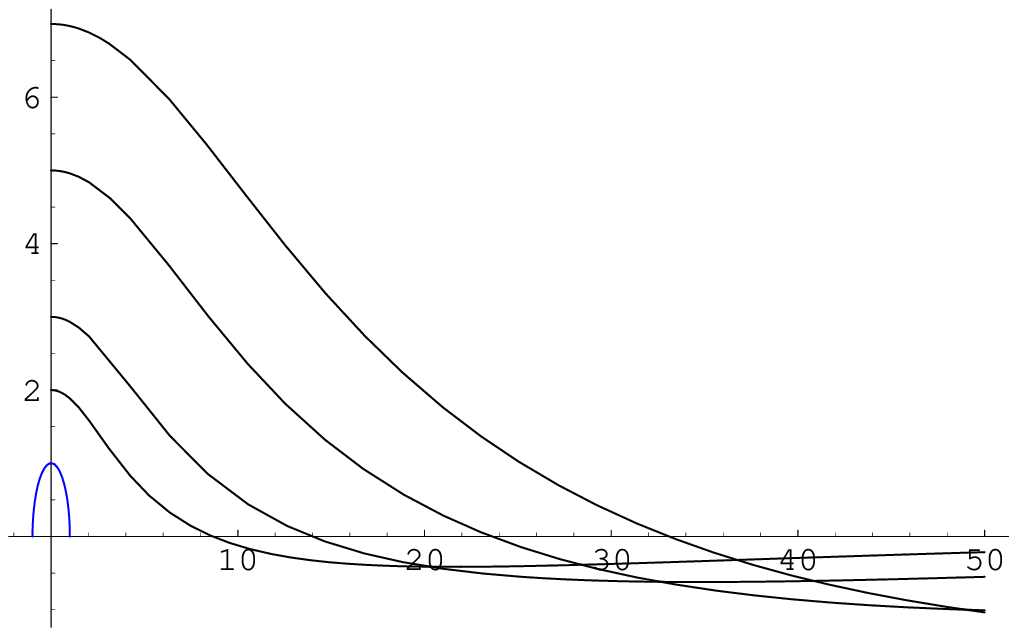}\label{fig:d7-01}}
\caption{ D7 brane embedding with different $y_0$ in the case of $Q=0$.  Blue circle denote minimal sphere in thermal $AdS_5$:  (a)  When D7 wraps compactified spatial direction.  (b)  When D7 does not wraps compactified spatial direction.
}
\end{center}
\end{figure}

One may ask what happen for D3 brane backgrounds with various probe branes. Naively
we expect similar behavior in this case. Indeed for D3 brane background with D5  probe brane,
completely parallel results are obtained.
However, for D3/D7 system with one direction compactified, the results are
not as expected.  For the deconfining black hole background we do not have any baryon vertex configuration as expected.
For the confining D3 background, there are two classes of  embedding of D7 brane
depending on whether D7 wraps  the compact direction or not. In case D7 is wrapping the compact direction,
the D7 does not get effectively repulsive force from the confining D3 background.
In fact  as far as D7 embedding is concerned, there is no difference in equation of motion between the black hole background and confining background, hence for small enough current quark mass, the
D7 embedding should fall into the singularity, an impossible result.  Therefore we do not expect that
such D7 embedding is allowed. See FIG. \ref{fig:d7I-01}.
If D7 is not wrapping the compact direction, numerical study shows that no stable D7 configuration with
finite current quark mass is allowed. See FIG. \ref{fig:d7-01}.
We leave the  description of some details to the appendix.

\section{Discussion}
We considered the issue of presence of Minkowski brane in the context of  baryon mass in the medium.
Although no analytic proof was made, our numerical investigation shows strong evidence of the absence of the
baryon vertex operator   in the black hole background.
In the dual picture, this corresponds to the absence of the baryonic states in the N=2 Yang Mills system with finite quark mass system. On the other hand, we found the existence of such states in the confining background like D3/D7 and D4/D6 double wick-rotation.
Such picture is consistent with the naive expectations from the yang-Mills side that baryons exist only in the confining phase.
%We studied  existence of baryon in a various background and the dependence of the baryon mass on the baryon density. Here we find that for large enough current quark mass, baryon mass is decreasing.

  It is  interesting to classify all the solution of the DBI action with source.
   The spiky solutions ( in deconfining background) represent quarks,  while the cusp solutions (in confining)
   represent the baryons. Now,
     DBI action allows solutions which is regular everywhere. How should we interpret this solution in string theory setting? In fact, the very motivation of the Born-Infeld action is to regularize electromagnetic singularity at the point charge.  The answer to this question is following: in the presence of dense medium, there are two scales in the system; one is the string scale controlling the scale of the spike and the other is the inter-particle distance.
     There are many order of magnitude difference between these scale. Therefore, the structure of one scale should be
     neglected when we treat that of the other scale. That is, when we focus on the physics of density, the
    spiky structure  should be neglected.  If we neglect the latter, then we get the smooth brane without singularity
    or cusp. This is the interpretation of the regular solution identified as  the Minkowski embedding in \cite{NSSY}.
    However, we should not forget that there are invisible strings  attached to the probe branes to give charge sources and their energy should be added to the total energy even though the shape of the  regulated brane is smooth.
    In fact, if we add the string mass to Minkowski brane solution, the sum is bigger than that of the the black hole
    embedding, which means that none of the Minkowski embeddings  in the quark phase are physically realized.

Notice that for the baryon phase  also,  the D5 mass should be always included as far as the baryonic states exist.
In QCD, most of the mass of the baryons are coming from the gluons instead of current quark mass.
This has an  analogue in hQCD:  for the
 baryon vertex in gravity dual, the string length is always minimized and effectively strings are
 replaced by the deformed probe brane. This suggests that in hQCD, the quarks are melt away in confining
 phase and  all the mass of the baryon is coming from the coherent gluons which is dual to the compact D-branes
 in the bulk.

 {\bf Note added}:
 After finishing this work, we received a relate work \cite{Ghoroku} where the the baryon vertex
 in the absence of the probe branes was  discussed.

%\section{D3/D5 system}
\appendix

\section{D3 brane - confining background}
We start usual Euclidean D3 black hole geometry.
\begin{eqnarray}
ds^{2}
&=&\frac{U^{2}}{R^{2}}\left(dt^{2}+d\vec{x}^{2}+f(U) d\t^{2}\right)
+R^{2}\left(\frac{dU^{2}}{f(U) U^{2}}+d\Omega_{5}^{2}\right),
\label{adsm}
%\\
%*dC_{(4)}&=&16\pi l_{s}^{4}N_{c}\epsilon_{(5)},
%\label{RR-flux}
\end{eqnarray}
where $R^{4}=2\lambda l_{s}^{4}=4\pi g_s N_c l_s^4$ and $f(U) =1-\left({U_{KK}}/{U}\right)^{4}$.
% and $C_{(4)}$ is the background RR 4-form field.
Here, $\lambda=g_{YM}^{2}N_{c}$ is the 't Hooft coupling of the YM theory.
$\t$ direction is compactified with period
\begin{eqnarray}
\frac{1}{\beta_{\t}}=\frac{U_{KK}}{\pi R^{2}}=\frac{M_{KK}}{2\pi},
\end{eqnarray}
where $M_{KK}$ is Kaluza-Klein mass.
Introducing a dimensionless coordinate $\xi$ defined by
$ {d\xi^2}/{\xi^2}= {dU^2}/({U^2f(U)})$, the bulk geometry becomes
\be\label{bgmetric}
ds^2 = \frac{U^2}{R^2}\left(dt^2 +f d\tau^2 +d\vec{x}^2 \right)
+\frac{R^2}{\xi^2}\left(d\xi^{2}+\xi^{2}d\Omega_{5}^{2}\right),
\ee
where
$U$ and $\xi$ are related by
$ {U^{2}}/{U_{KK}^2}=\frac{1}{2} (\xi^{2}+{1}/{\xi^{2}})\,$ and $f = ( {1- \xi ^{4}})^2/({1+ \xi^{4}} )^{2}.$

 From now on, we will use the metric (\ref{bgmetric}) as a background metric
with baryon vertex and probe brane.

\subsection{Baryon vertex - D5 brane}
The brane configuration which corresponds to the baryon vertex can be understood as D5 brane wrapping compact $S^5$
direction. The background metric (\ref{bgmetric}) can be written as
\be
ds^2 = \frac{U^2}{R^2}\left(dt^2 +f d\tau^2 +d\vec{x}^2 \right)
+\frac{R^2}{\xi^2}d\xi^2 +R^2\left(d\theta^2 +\sin^2\theta d\Omega_{4}^{2}\right),
\ee
We assume that only $\xi$ depends on $\theta$ and $F_{t\theta}\ne 0$. The induced metric on D5 brane is
\be\label{d5met}
ds_{D5}^2 = \frac{U^2}{R^2}dt^2 + R^2\left(1+\frac{\xi'^2}{\xi^2}\right)d\theta^2 +R^2\sin^2\theta d\Omega_{4}^2.
\ee
The DBI action for single D5 brane with $N_c$ fundamental string can be written as similar to \cite{Callan:1999zf}
\bea
S_{D5} &=& -\mu_5 \int \sqrt{{\rm det}(g+2\pi \a' F)}+\mu_5 \int A_{(1)}\wedge G_{(5)} \cr\cr
&=&\t_5 \int dtd\theta \sin^4\theta\left[-\sqrt{\omega_+ (\xi^2 +\xi'^2)
-\tilde{F}^2} +4\tilde{A}\right],
\eea
where
\be
\t_5 = \frac{\mu_5 \O_4 R^4 U_{KK}}{\sqrt{2}},~~~~
\tilde{F} = \frac{2\sqrt{2}\pi \a' F_{t\theta}}{U_{KK}},~~~~~~\tilde{A_t}=\frac{\sqrt{2}\cdot 2\pi \a'}{U_{KK}} A_t.
\ee
and $\xi' =\partial \xi/\partial \theta$.\par
The displacement can be defined as follows;
\bea
\frac{\partial {\cal L}_{D5}}{\partial \tilde{F}}= \frac{\sin^4\theta \tilde{F}}{\sqrt{\omega_+ (\xi^2 +\xi'^2)
-\tilde{F}^2}} \equiv& -D(\theta).
\eea
From the equation of motion for gauge field, we can obtain
\be
D(\theta)=\left[\frac{3}{2}(\nu\pi-\theta)+\frac{3}{2}\sin\theta\cos\theta+\sin^3\theta\cos\theta\right],
\ee
where $\nu$ determines the number of fundamental sting( $\nu N_c$ strings are at south pole and $(1-\nu)N_c$ strings
are at north pole, we set $\nu =0$)\par
By performing Legendre transformation, we can get `Hamiltonian'
\bea\label{d5h}
{\cal H}_{D5} &=&\tilde{F}\frac{\partial {\cal L}_{D5}}{\partial \tilde{F}}-{\cal L}_{D5}\cr\cr
&=&\t_5 \int d\theta \sqrt{\omega_+ (\xi^2 +\xi'^2)}\sqrt{D(\theta)^2+ \sin^8\theta}
\eea
Solutions of equation of motion for above Hamiltonian are characterized by initial value of $\xi_0$ at $\theta=0$(we set
$\xi'(\theta=0)=0$ as a initial condition. The solutions for different $\xi_0$ are qualitatively same as FIG. \ref{fig:baryon-d4}.

The force at the cusp of D5 brane tension is
\bea\label{force-d5}
F_{D5} &=& \frac{\partial{{\cal H}}}{\partial U_c} \Bigg|_{fix~other~values} \cr\cr
&=& N_c T_F \left(\frac{1+\xi_c^{-4}}{1-\xi_c^{-4}}\right)
 \frac{\xi'_c }{\sqrt{\xi_c^2 +\xi_c'^2}},
\eea
where $T_F$ is tension of fundamental string.
The force is always smaller than the tension of fundamental string same as baryon D4 brane case.
Therefore, we have to find stable configuration with probe D-branes which make hole system to be stable.

\subsection{Probe D5 brane}
We put $N_f$ number of D5 branes as probe flavor brane. The background metric (\ref{bgmetric})
can be written as
\be
ds^2 = \frac{U^2}{R^2}\left(dt^2 +f d\tau^2 +dx_2^2 +dx_3^2 \right)
+\frac{R^2}{\xi^2}\left(dy^2+d\r^2 +\r^{2}d\Omega_{2}^{2}+dz^2 +z^2 d\phi^2 \right),
\ee
where D5 brane world volume coordinates are $(t,x_2,x_3,\r,\t_{\a})$.
We also assume that the only coordinate $y$ depends on $\r$ and focus on $z=\phi=0$ solution.
The induced metric on D5 brane is
\be
ds_{D5}^2 = \frac{U^2}{R^2}\left(dt^2 + dx_2^2 +dx_3^2 \right) +\frac{R^2}{\xi^2}\left[(1+\dot{y}^2)d\r^2
+\r^2 d\O_2^2 \right]
\ee
with non-zero electric field on it ($F_{\r t} \ne 0$) and $\dot{y}=\partial y/\partial\r$. \par
DBI action of probe D5 brane is
\be
\hat{S}_{D5}
= -\hat{\t}_5\int dtd\r \r^2 \o_+\sqrt{\o_+(1+\dot{y}^2)-\tilde{F}^2}\equiv \int dt \hat{\cal L}_{D5}
\ee
where
\bea
\hat{\t}_5 =\frac{1}{2\sqrt{2}}N_f \mu_5 \O_2 V_2  U_{KK}^3  ,\;\;\;
\tilde{F}=\frac{2\sqrt{2}\pi\a'F_{\r t}}{U_{KK}}.
\eea
We define dimensionless value $\tilde{Q}$ from equation of motion for $\tilde{F}$ is
\be
\frac{\partial \hat{S}_{D5}}{\partial \tilde{F}}=\frac{\r^3\o_+\tilde{F}}
{\sqrt{\o_+(1+\dot{y}^2)-\tilde{F}}}
\equiv \tilde{Q}
\ee
The Legendre transformed `Hamiltonian' from is
\be\label{d5hh}
\hat{\cal H}_{D5}
%&=& \tilde{F}\frac{\partial \hat{{\cal L}}_{D5}}{\partial \tilde{F}}-\hat{{\cal L}}_{D5} \cr\cr
=\hat{\t}_5\int d\r \sqrt{\o_+\left(\tilde{Q^2} +\r^4\o_+^2\right)} \sqrt{1+\dot{y}^2}
%&=&\hat{\t}_5 \int d\r V(\r) \sqrt{1+\dot{y}^2},
\ee
with
\be
\tilde{Q}=\frac{U_{KK} Q}{2\pi\a' \sqrt{2}\hat{\t}_5},
\ee
where $Q$ is total number of quarks.\par
In $\tilde{Q}=0$ case, the solutions of probe D5 brane embedding are same as FIG. \ref{fig:d6-01}.
In $\tilde{Q}\ne 0$ case, we can impose force balance condition
\be\label{force-balance}
\hat{F}_{D5}= \frac{Q}{N_c} F_{D5}
\ee
with following constraint
\be
\dot{y}_c =\frac{\xi_c'}{y_c}.
\ee
The D-brane configurations with and without FBC and the density dependence
of baryon mass are qualitatively same as FIG. \ref{d6forcebal} and FIG. \ref{fig:md4s-01}.

\subsection{Probe D7 brane}
In this section, we consider D7 brane configuration used in \cite{NSSY,NSSY2}. In this case the background metric
(\ref{bgmetric}) can be written as
\be
ds^2 = \frac{U^2}{R^2}\left(dt^2 +f d\tau^2 +dx_2^2 +dx_3^2 \right)
+\frac{R^2}{\xi^2}\left(dy^2+d\r^2 +\r^{2}d\Omega_{3}^{2}+y^2 d\phi^2\right),
\ee
where $\xi^2 =y^2 +\r^2$. D7 brane wraps compactified direction $\t$.
We also assume that the only coordinate $y$ depends on $\r$. The induced metric on D7 brane is
\be
ds_{D7}^2 = \frac{U^2}{R^2}\left(dt^2 + f d\t^2 +dx_2^2 +dx_3^2 \right) +\frac{R^2}{\xi^2}\left[(1+\dot{y}^2)d\r^2
+\r^2 d\O_3^2 \right]
\ee
with non-zero electric field on it ($F_{\r t} \ne 0$) and $\dot{y}=\partial y/\partial\r$. \par
DBI action of D7 brane is
\be
S_{D7}
=-\t_7 \int dtd\r \r^3\o_-\sqrt{\o_+}
\sqrt{\o_+(1+\dot{y}^2) -\tilde{F}^2}
\ee
where,
\be
\t_7 = \frac{1}{4}N_f \mu_7 \O_3 \beta_{\t} V_2 U_0^2 ,~~~~~
\tilde{F} = \frac{2\sqrt{2}\pi\a'F_{\r t}}{U_0}.
\ee
Equation of motion for $\tilde{F}$ is
\be
\frac{\partial {\cal L}_{D7}}{\partial \tilde{F}}=\frac{\r^3\o_-\o_+^{1/2}\tilde{F}}{\sqrt{\o_+(1+\dot{y}^2)-\tilde{F}}}
\equiv  \tilde{Q}
\ee
We can get a Hamiltonian from Legendre transformation,
\be\label{d7Ih}
{\cal H}_{D7}
=\t_7\int d\r \sqrt{\o_+\left(Q^2 +\r^6 \o_-^2\o_+\right)} \sqrt{1+\dot{y}^2}.
\ee
In the case of $Q=0$, the Hamiltonian (\ref{d7Ih}) is same as the Hamiltonian for black hole background.
However, in confining background, probe brane cannot end on the singularity. It means that
probe brane cannot cover low $y_{\infty}$ region in $Q=0$ case.\par

\subsection{Probe - D7 brane (II)}
 Here we consider the probe D7 brane does not contain compact direction.
 The background metric (\ref{bgmetric}) can be written as
\be
ds^2 = \frac{U^2}{R^2}\left(dt^2 +f d\tau^2 +dx_2^2 +dx_3^2 \right)
+\frac{R^2}{\xi^2}\left(dy^2+d\r^2 +\r^{2}d\Omega_{4}^{2}\right),
\ee
where $\xi^2 =y^2 +\r^2$.
We also assume that   only  $y$ depends on $\r$. The induced metric on D7 brane is
\be
ds_{D7}^2 = \frac{U^2}{R^2}\left(dt^2 + dx_2^2 +dx_3^2 \right) +\frac{R^2}{\xi^2}\left[(1+\dot{y}^2)d\r^2
+\r^2 d\O_4^2 \right]
\ee
with non-zero electric field on it ($F_{\r t} \ne 0$) and $\dot{y}=\partial y/\partial\r$. \par
DBI action of D7 brane is
\bea
S_{D7}
&=&-\t_7 \int dtd\r \frac{\r^4 \o_+}{\xi^2}
\sqrt{\o_+(1+\dot{y}^2) -\tilde{F}^2} ,
\eea
where,
\be
\t_7 =\frac{1}{2\sqrt{2}}N_f \mu_7 \O_4 V_2 R^2 U_0^3, ~~~~~~~
\tilde{F} = \frac{2\sqrt{2}\pi\a'F_{\r t}}{U_0}
\ee
Equation of motion for $\tilde{F}$ is
\be
\frac{\partial {\cal L}_{D7}}{\partial \tilde{F}}=\frac{\r^4\o_+\tilde{F}}{\xi^2\sqrt{\o_+(1+\dot{y}^2)-\tilde{F}}}
\equiv  \tilde{Q}
\ee
We can get a Hamiltonian from Legendre transformation,
\be\label{d7h}
{\cal H}_{D7}
%&=& \tilde{F}\frac{\partial {\cal L}_{D7}}{\partial \tilde{F}}-{\cal L}_{D7} \cr\cr
=\t_7\int d\r \sqrt{\o_+\left(Q^2 +\frac{\r^8\o_+^2}{\xi^4}\right)} \sqrt{1+\dot{y}^2}
%&=&\t_7 \int d\r V(\r) \sqrt{1+\dot{y}^2}.
\ee
As we discussed before, the solution of equation of motion for above Hamiltonian gives funny behavior even $Q=0$ case.
D7 brane does not goes to flat brane as $y_0$ increase. Moreover, values of $y_\infty$ are very small for every $y_0$
even for a very large value of $y_0$. It means that the value of $y_\infty$ cannot cover whole $y$ axes.
And therefore we do not consider this case seriously.

\vskip 1.5cm

\noindent{\bf \large Acknowledgement}\\
We want to thank Shin Nakamura and Mannque Rho for useful discussion and comments.
We also like to thank Jonathan Shock for his  careful reading and correcting an error in eq.11
of the first version.
This work is supported  by KOSEF Grant  R01-2007-000-10214-0.
The work of SJS is also supported by the Korea Research Foundation Grant funded by the Korean Government
(MOEHRD, Basic Research Promotion Fund) (KRF-2007-314-C00052)
and by  the SRC Program of the KOSEF through the CQUEST  grant R11-2005-021.

\end{document}